\documentclass[11pt,a4paper]{article}
\usepackage[hyperref]{acl2021}
\usepackage{times}
\usepackage{latexsym}

\usepackage[T1]{fontenc}

\usepackage{microtype}

\aclfinalcopy % Uncomment this line for the final submission
\usepackage{microtype}
\usepackage{booktabs}
\usepackage{multirow}
\usepackage{latexsym}
\usepackage{float}
\restylefloat{table}

\usepackage{amsmath}
\usepackage{amssymb}

\usepackage{url,graphicx}
\graphicspath{ {fig/} }

\title{ASR Adaptation for E-commerce Chatbots using Cross-Utterance Context and Multi-Task Language Modeling}

\author{Ashish Shenoy \quad Sravan Bodapati \quad Katrin Kirchhoff\\
  Amazon AWS AI, USA\\
  {\tt \{ashenoy, sravanb, katrinki\}@amazon.com} \\
}

\date{}

\begin{document}
\maketitle
\begin{abstract}
Automatic Speech Recognition (ASR) robustness toward slot entities are critical in e-commerce voice assistants that involve monetary transactions and purchases. Along with effective domain adaptation, it is intuitive that cross utterance contextual cues play an important role in disambiguating domain specific content words from speech. In this paper, we investigate various techniques to improve contextualization, content word robustness and domain adaptation of a Transformer-XL neural language model (NLM) to rescore ASR N-best hypotheses. To improve contextualization, we utilize turn level dialogue acts along with cross utterance context carry over. Additionally, to adapt our domain-general NLM towards e-commerce on-the-fly, we use embeddings derived from a finetuned masked LM on in-domain data. Finally, to improve robustness towards in-domain content words, we propose a multi-task model that can jointly perform content word detection and language modeling tasks. Compared to a non-contextual LSTM LM baseline, our best performing NLM rescorer results in a content WER reduction of 19.2\% on e-commerce audio test set and a slot labeling F1 improvement of 6.4\%.

\end{abstract}

\section{Introduction}

Task-oriented conversations in voice chatbots deployed for e-commerce usecases such as shopping \cite{maarek:2018}, browsing catalog, scheduling deliveries or ordering food are predominantly short-form audios. Moreover, these dialogues are restricted to a narrow range of multi-turn interactions that involve accomplishing a specific task \cite{mari:20}. The back and forth between a user and the chatbots are key to reliably capture the user intent and slot entities referenced in the spoken utterances. As shown in previous works \cite{irie:19, partha:19, sun2021transformer}, rather than decoding each utterance independently, there can be benefit in decoding these utterances based on context from previous turns. In the case of grocery shopping for example, knowing that the context is "what kind of laundry detergent?" should help in disambiguating "pods" from "pause". Another common aspect in e-commerce chatbots is that the speech patterns differ among sub-categories of usecases (Eg. shopping clothes vs ordering fast food). Hence, some chatbot systems allow users to provide pre-defined grammars or sample utterances that are specific for their usecase \cite{aggandhe:18}. These user provided grammars are then predominantly used to perform domain adaptation on an n-gram language model. Recently \cite{shenoy2021whats} showed that these can be leveraged to bias a Transformer-XL (TXL) LM rescorer on-the-fly.

While there has been extensive previous work on improving contextualization of TXL LM using historical context, none of the approaches utilize signals from a natural language understanding (NLU) component such as turn level dialogue acts. This paper investigates how to utilize dialogue acts along with user provided speech patterns to adapt a domain-general TXL LM towards different e-commerce usecases on-the-fly. We also propose a novel multi-task architecture for TXL, where the model jointly learns to perform domain specific slot detection and LM tasks. We use perplexity (PPL) and word error rate (WER) as our evaluation metrics. We also evaluate on downstream NLU metrics such as intent classification (IC) F1 and slot labeling (SL) F1 to capture the success of these conversations.
The overall contributions of this work can be summarized as follows :
\begin{itemize}
    \item We show that a TXL model that utilizes turn level dialogue act information along with long span context helps with contextualiziation and improves WER and IC F1 in e-commerce chatbots.
    \item To improve robustness towards e-commerce domain specifc slot entities, we propose a novel TXL architecture that is jointly trained on slot detection and LM tasks which significantly improves content WERR and SL F1.
    \item We show that adapting the NLM towards user provided speech patterns by using BERT on domain specific text is an efficient and effective method to perform on-the-fly adaptation of a domain-general NLM towards e-commerce utterances. 
\end{itemize}

\section{Related Work}
Incorporating cross utterance context has been well explored with both recurrent and non-recurrent NLMs. With LSTM NLMs, long span context is usually propogated without resetting hidden states across sentences or using longer sequence lengths \cite{ms:2018, irie:19, khandelwal-etal-2018-sharp,partha:19}. In \cite{xiong-etal-2018-session}, along with longer history, information about turn taking and speaker overlap is used to improve contextualization in human to human conversations. With transformer architecture based on self attention \cite{vaswani:17} \cite{dai-etal-2019-transformer} showed that by utilizing segment wise recurrence Transformer-XL (TXL) \cite{dai-etal-2019-transformer} is able to effectively leverage long span context while decoding. More recently, improving contextualization of the TXL models included adding a LSTM fusion layer to complement the advantages of recurrent with non-recurrent models \cite{sun2021transformer}. \cite{shenoy2021whats} incorporated a non-finetuned masked LM fusion in order to make the domain adaptation of TXL models quick and on-the-fly using embeddings derived from customer provided data and incorporated dialogue acts but only with an LSTM based LM. While \cite{sunkaral:2020} tried to fuse multi-model features into a seq-to-seq LSTM based network. In \cite{sharma-2020-improving} cross utterance context was effectively used to perform better intent classification with e-commerce voice assistants.

For domain adaptation, previous techniques explored include using an explicit topic vector as classified by a separate domain classifier and incorporating a neural cache \cite{mikolov:12,keli:2018,anirudh:18,Chen:2015}. \cite{irie18:radmm} used a mixture of domain experts which are dynamically interpolated. It is also shown in \cite{Liu:2020}, that using a hybrid pointer network over contextual metadata can also help in transcribing long form social media audio. 
Joint learning NLU tasks such as intent detection and slot filling have been explored with RNN based LMs in \cite{liu-lane-2016-joint} and more recently in \cite{Milind:20}, where they show that a jointly trained model consisting of both ASR and NLU tasks interfaced with a neural network based interface helps incorporate semantic information from NLU and improves ASR that comprises a LSTM based NLM. In \cite{yang:20} tried to incorporate joint slot and intent detection into a LSTM based rescorer with a goal of improving accuracy on rare words in an end-to-end ASR system. 

However, none of the previous work utilize dialogue acts with a non-recurrent based LM such as Transformer-XL nor optimize towards improving robustness of in-domain slot entities. In this paper we experiment and study the impact of utilizing dialogue acts along with a masked language model fusion to improve contextualization and domain adaptation. Additionally, we also propose a novel multi-task architecture with TXL LM that improves the robustness towards in-domain slot entity detection.

\section{Approach}
A standard language model in an ASR system computes a probability distribution over a sequence of words $W = w_0,...,w_N$ auto-regressively as:
\begin{equation}\label{eq:1}
p(W) =\prod_{i=1}^N{p(w_i|w_1, w_2,...,w_{i-1})}
\end{equation}

In our experiments, along with historical context, we condition the LM on additional contextual metadata such as dialogue acts : 
\begin{equation}\label{eq:1}
p(W) =\prod_{i=1}^N{p(w_i|w_1, w_2,...,w_{i-1}, c_1, c_2,..., c_k)}
\end{equation}

Where $c_1, c_2,...c_k$ are the turn based lexical representation of the contextual metadata. For baseline, we use a standard LSTM LM as summarized below : 
\begin{equation}\label{eq:3}
\begin{aligned}
&embed_i = E^T_{ke}w_{i-1} \\
& c_i, h_i = LSTM(h_{i-1}, c_{i-1}, embed_i) \\
&p(w_i|w_{<i}) = Softmax(W^T_{ho}h_i) \\
\end{aligned}
\end{equation}
where $embed_i$ is a fixed size lower dimensional word embedding and the LSTM outputs are projected to word level outputs using $W^T_{ho}$. A $Softmax$ layer converts the word level outputs into final word level probabilities.

\begin{figure}[!t]
    \includegraphics[width=0.9\linewidth]{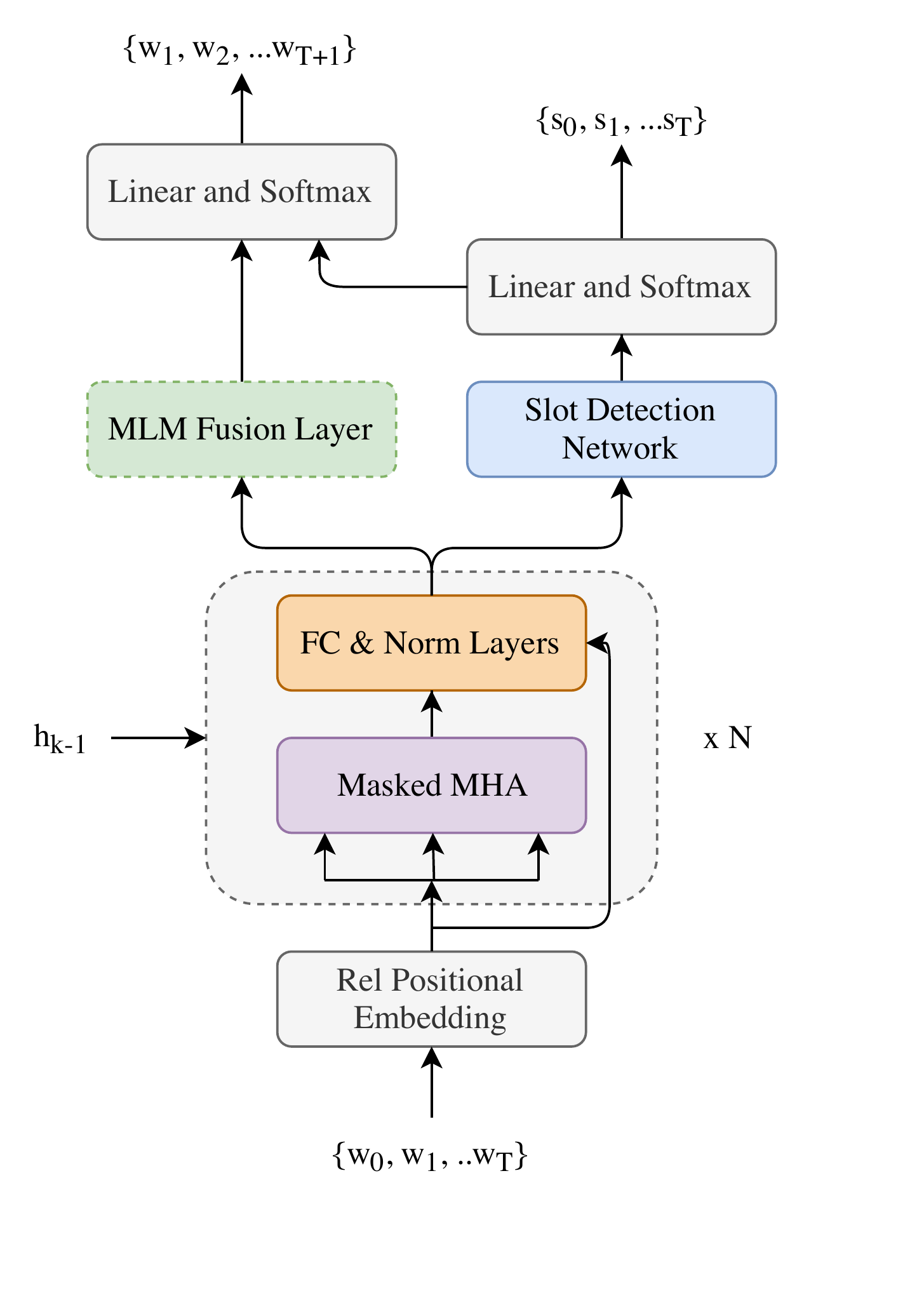}
    \caption{Transformer-XL language model architecture jointly trained with slot detection task with an optional MLM fusion layer}
    \label{fig:txl_joint}
\end{figure}

\begin{figure}
    \includegraphics[width=1.0\linewidth]{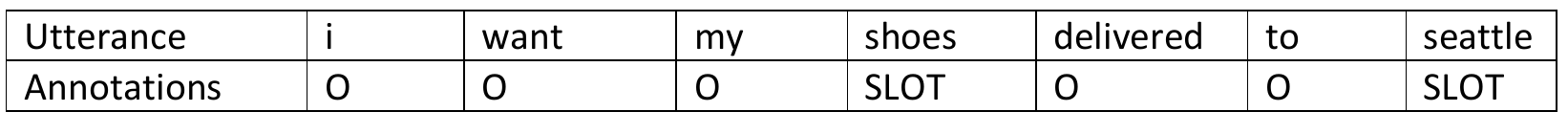}
    \caption{Example utterance with slots annotated}
    \label{fig:eg_utt}
\end{figure}

\subsection{Transformer-XL based NLM}
Although recurrent language models help in modeling long range dependencies to certain extent, they still suffer from the fuzzy far away problem \cite{khandelwal-etal-2018-sharp}. Vanilla transformer LMs on the other hand use fixed segment lengths which leads to context fragmentation. To address these limitations and model long range dependencies, TXL models add segment-level recurrence and use a relative positional encoding scheme \cite{dai-etal-2019-transformer}. Hence we choose to use a TXL LM directly. The cached hidden representations from previous segments helps contextual information flow across segment boundaries.  If $s_k = [w_{k,1}, ..., w_{k,T}]$ and $s_{k+1} = [x_{k+1,1},...,x_{k+1,T}]$ are two consecutive segments of length $T$ and $h^n_k$ is the $n$-th layer hidden state produced for the $k$-th segment $s_k$, then, the $n$-th layer hidden state for segment $s_{k+1}$ is produced as follows:
\begin{equation}
\begin{aligned}
\tilde{h}^{n-1}_{k+1} = [SG(h_k^{n-1}) \circ h_{k+1}^{n-1}] \\
q^n_{k+1}, k^n_{k+1}, v^n_{k+1} = \tilde{h}_{k+1}^{n-1}{W_q}^\intercal \\
h^n_{k+1} = TL(q_{k+1}^n, k^n_{k+1}, v^n_{k+1})
\end{aligned}
\end{equation}
where $SG(.)$ stands for stop gradient and $TL$ stands for Transformer Layer.
To carry over context from previous turns, we train and evaluate the model by concatenating all the turns, including the bot responses, in a single conversation session. The model is trained with a cross entropy objective as defined below :

\begin{equation}
\begin{aligned}
    \mathbf{L_{LM}} = - \frac{1}{T} \left [\sum_{i=1}^T \mathit{log}(P(w_{i}\mid w_{<i}, s_{<i}))\right ] \\
\end{aligned}
\end{equation}

During inference time, we cache a fixed length hidden representation from previous segments. We also use the generated bot responses to perform a forward pass and carry over the context to the next user turn.

\subsection{Slot detection and language modeling multi-task learning}
To make the our domain-general model robust to e-commerce specific slot entities, we propose a multi-task learning approach to training the TXL LM. We train our models on both LM and slot detection tasks. Similar to slot filling, slot detection is a sequence classification task that involves predicting if a word, $w_i$ at time step $i$ is a domain specific slot entity. We use a separate slot detection network, consisting of a simple multi-layer perceptron, and use the final layer hidden representation from the TXL network as inputs to the network. Figure \ref{fig:eg_utt} shows an example utterance with the slot annotations. Formally, let $s = (s_0, s_1, ..., s_T )$ be the slot label sequence, corresponding to a word sequence $w = {w_0, w_1, ...., w_T}$ in the $k$-th segment. We model the slot label output $s_t$ as a conditional distribution over input word sequence up to time step $t$, $w_{\leq t}$ similar to \cite{liu-lane-2016-joint} :

\begin{equation}
\begin{aligned}
h_k^n = TL(q_{k}^n, k^n_{k}, v^n_{k}) \\
p(s_t|w_{\leq t}) = SlotLabelDist(h_t^n)
\end{aligned}
\end{equation}

We use a cross-entropy training objective for the slot detection task as below :  

\begin{equation}
\begin{aligned}
    \mathbf{L_{SD}} = - \frac{1}{T} \left [\sum_{i=1}^T \mathit{log}(P(s_i\mid w_{\leq i}))\right ] \\
\end{aligned}
\end{equation}

To incorporate this semantic information about the word from previous time step into the NLM, we use the logits from the slot detection network to condition the probability distribution of the next word in the sequence as shown in Figure \ref{fig:txl_joint}.

The total loss is then computed using a linear combination of LM and slot detection losses:

\begin{equation}
\begin{aligned}
    \mathbf{L_{total}} = \mathbf{L_{LM}}+\alpha_{SD}\mathbf{L_{SD}}
\end{aligned}
\end{equation}
where $\alpha_{SD}$ is the weight for the slot detection loss. \\

\subsection{Transformer-XL LM conditioning on dialogue acts}
Dialogue acts (DA) in a conversation represent the intention of an utterance and is intended towards capturing the action that an agent is trying to accomplish \cite{john_austin:75}. An example conversation snippet with DA is shown in Table \ref{tab:eg_da}. DA classification is typically performed in a separate component that is part of a downstream NLU system and consumes the outputs generated by ASR. The classified DA is an important contextual signal that provides hints about the type of speech pattern that can be expected in the next turn. We utilize these signals to train our TXL models. Specifically, we augment the training data with the dialogue act information prefixed to the user turns and surround them with explicit <dialogue\_act> tags. The expectation is that the TXL LM learns the usage patterns associated with different dialogue acts and this information should help narrow down the search space for the model to content words relevant to the current dialogue context.

\begin{table}
\centering
\footnotesize
    \small
    \begin{tabular}{lll}
    \toprule
    \textbf{Actor} & \textbf{Utterance} & \textbf{Dialogue Act}\\
    \midrule
    Bot & how can i help you today & general-welcome \\
    User & hi i want to track my \\
    & online shopping order & inform-intent \\
    Bot & sure! what is the order \\
    & number? & request \\
    User & my order number is abcdef & inform \\
    Bot & your order is scheduled to be \\
    & delivered tomorrow & inform \\
    User & thanks & thank-you \\
    Bot & do you need help with \\
    & anything else? & req-more \\
    \bottomrule
    \end{tabular}
\caption{A sample user bot conversation snippet showing example dialogue acts.}
\label{tab:eg_da}
\end{table}

\begin{table*}[th!]
  \centering
  \small
  \resizebox{\textwidth}{!}{%
    \begin{tabular}{llrrrrrrrr}
        \toprule
         &\textbf{Model} & \multicolumn{4}{c}{\textbf{Retail}} & \multicolumn{4}{c}{\textbf{Fastfood}} \\
              & & \textbf{CWERR} & \textbf{IC F1} & \textbf{SL F1} & \textbf{p-value} & \textbf{CWERR} & \textbf{IC F1} & \textbf{SL F1} & \textbf{p-value} \\
        \midrule
        1 & Non-contextual LSTM & -- & -- & -- & -- & -- & --  \\
        \midrule
        2 & TXL & 1.0\% & 0.5\% & 0.4\% & 0.083 & 12.3\% & 0.4\% & 0.9\% & \textbf{0.048} \\
        \midrule
        3 & \hspace{0.5em}+ Dialogue Acts (DA) & 1.2\% & 1.2\% & 1.3\% & 0.057 & 14.4\% & 1.0\% & 1.2\% & \textbf{0.041} \\
        \midrule
        4 & \hspace{0.5em}+ Joint Slot Detection (SD) & 4.3\% & 2.0\% & 3.3\% & \textbf{0.046} & 16.3\% & 0.9\% & 2.1\% & \textbf{0.015}\\
        5 & \hspace{0.5em}+ Joint SD + DA & 8.6\% & 2.1\% & 3.3\% & \textbf{0.048} & 17.3\% & 1.3\% & 2.7\% & \textbf{0.009} \\
        \midrule
        6 & \hspace{0.5em}+ BERT Fusion & 6.4\% & 2.8\% & 2.3\% & \textbf{0.030} & 18.2\% & 1.8\% & 4.8\% & \textbf{0.004}  \\
        7 & \hspace{0.5em}+ BERT Fusion + DA & 9.6\% & 2.9\% & 2.7\% & \textbf{0.023} & 19.2\% & 1.7\% & 4.8\%  & \textbf{0.004} \\
        \midrule
        8 & \hspace{0.5em}+ Joint SD + DA + BERT Fusion & 11.8\% & 3.8\% & 4.3\% & \textbf{0.037} & 19.2\% & 2.1\% & 6.4\% & \textbf{0.002}\\
     \bottomrule
    \end{tabular}
    }
    \caption{ASR and NLU improvements on two e-commerce sub-domains : Retail and Fastfood. CWERR - Content Word Error Reduction, IC F1 - Relative Intent Classification F1 Improvement, SL F1 - Relative Slot Labeling F1 Improvement, MPSSWE p-value test on WERR where significant improvements are in bold}
    \label{table:wer_report}
\end{table*}

\subsection{Domain adaptation using contextual semantic embeddings}
In production chatbots, it is common for bot developers to provide example speech patterns, in the form of sample sentences or explicit grammars, which can then be used to bias the n-gram language models in a ASR system \cite{aggandhe:18}. This pre-defined set of speech patterns is a useful source of contextual information that can be also used to bias NLMs as well. As demonstrated in \cite{shenoy2021whats}, pretrained masked language models (MLM) such as BERT, can be used to derive a fixed size semantic representation from this lexical information.
Large pretrained MLMs are gaining widespread popularity and are considered as powerful language learners \cite{Radford2019,brown2020language}. However, the sentence or document embeddings derived from such an MLM without finetuning on in-domain data is shown to be inferior in terms of the ability to capture semantic information that can be used in similarity related tasks \cite{reimers-gurevych-2019-sentence}. Instead of using the [CLS] vector to obtain sentence embeddings, in this paper we take the average of context embeddings from last two layers as these are shown to be consistently better than using [CLS] vector \cite{reimers-gurevych-2019-sentence, li-etal-2020-sentence}.

We use a simple fusion method as experimented in \cite{shenoy2021whats} where the hidden state from the last layer of the TXL decoder is concatenated with the BERT derived embedding. This is then followed by a single projection layer with a non-linear activation function $\sigma$, such as $sigmoid$. \\
\begin{equation}\label{eq:6}
g_t = \sigma(W[h_t^{TXL}; e^{MLM}] + b)
\end{equation} 

Where $h_t^{TXL}$ is the hidden state from the last transformer decoder and $e^{MLM}$ is the BERT derived embedding from in domain sample utterances.
The intuition here is that the model learns to associate the domain specific BERT derived embedding with the occurrences of jargon specific to that domain. Thus providing different BERT vectors derived from different domain texts should allow the model to adapt towards such domains on-the-fly.

\begin{table}[h!]
\small
\centering
\footnotesize
    \begin{tabular}{l|c|c}
    \toprule
    \textbf{Model} & \textbf{PPLR$_{gen}$} & \textbf{PPLR$_{ecom}$}\\
    \midrule
    TXL & --  \\
    \hspace{0.5em} + Dialogue Acts (DA) & 3.4\%  & 9.9\%\\
    \hspace{0.5em} + Joint Slot Detection (SD) & 8.5\%  & 11.4\%\\
    \hspace{0.5em} + BERT Fusion (BF) & 16.3\%  & 23.3\%\\
    \midrule
    \hspace{0.5em} + Joint SD + DA & 9.8\% & 13.0\%\\
    \hspace{0.5em} + BF + DA & 21.5\%  & 25.3\%\\
    \hspace{0.5em} + BF + DA + Joint SD & 21.0\%  & 25.8\%\\
    \bottomrule
    \end{tabular}
\caption{Relative perplexity reduction (PPLR) from the various TXL models on a general domain eval set (PPL$_{gen}$) and on e-commerce domain eval set (PPL$_{ecom}$).}
\label{table:ppl_report}
\end{table}

\section{Experimental Setup}
\subsection{Dataset}
We required task-oriented dialogue datasets with actor, dialogue acts and the slot entities annotated. Since no single dataset was large enough to train a reliable language model, we used a combination of Schema-Guided Dialogue Dataset \cite{Rastogi:20}, MultiWOZ 2.1 \cite{MultiWoz:19, multiwoz:18}, MultiDoGo \cite{multidogo:19} along with anonymized in-house datasets that belong to two e-commerce usecases : retail and fastfood delivery. The final LM training data consisted of ~260k training samples, ~56k validation and evaluation samples and around 9.9 million running words. We used a vocabulary of size 25k. We evaluated our models on anonymized in-house 8kHz close-talk audio. These audio comprised of task-oriented conversations with multiple speakers and acoustic conditions representative of real world usage and belonged to the same two usecases mentioned above. The average number of turns in the audio dataset was 5.

\subsection{ASR setup and NLM setup}
We used a hybrid ASR model comprising of a regular-frame-rate (RFR) model trained on cross-entropy loss, followed by sMBR \cite{Ghoshal13}. The first pass LM we used was a domain-general Kneser-Ney (KN) \cite{KN:95} smoothed 4-gram model estimated on a weighted mix of datasets spanning multiple domains. The final vocabulary size of the n-gram LM was 500k words. 
All our NLM rescorers used a 4-layer Tranformer-XL\footnote{https://github.com/kimiyoung/transformer-xl} decoder, each of size 512 with 4 attention heads. The input word embedding size was 512. We used a segment and memory length of 25. During model training we applied a dropout rate of 0.3 to both the slot detection network and TXL. For the slot detection layer we used a 3 layer MLP and used the final layer hidden representation from the TXL as the output.
To obtain the BERT embedding from in-domain speech patterns, we finetune huggingface\footnote{https://github.com/huggingface/transformers} pretrained BERT mode on the retail and fastfood text corpus. The derived BERT embedding size used was 768.
During inference, we extract n-best hypothesis with \verb|n<=50| from the lattice generated by the first pass ASR model. We rescored the n-best hypothesis by multiplying the acoustic score with the acoustic scale and adding it to the scores obtained from the TXL rescorer. We used a fixed $\alpha_{SD}$ of 0.8 for the slot detection loss.

\section{Results and Discussion}
% DAs improve on intent detection F1 but not so much on actual content words
% Adding joint slot detection loss improves WER on content words and SL F1 over baseline
% BERT Fusion helps the most towards context adaptation with both IC and SL increasing on both domains
% Combining all of the methods amplify the gains
Table \ref{table:ppl_report} summarizes the relative perplexity reductions (PPLR).  Since we are optimizing our models to improve on the e-commerce domain specific content words we directly report the relative content word error rate reductions (CWERR) in Table \ref{table:wer_report} along with the relative impact on the downstream NLU tasks of IC and SL. For computing CWERR, we remove all the stop words comprising of commonly used function words, such as conjunctions and prepositions from the transcriptions and evaluate only on content words. We also report statistical significance of our CWER improvements using matched pairs sentence segment word error test (MPSSWE).
All the WER numbers are relative to a non-contextual LSTM baseline. The gap in the performance between the two domains we tested on is reflective of the underlying training corpus distribution, which has more text belonging to the fastfood domain.

\textbf{Perplexity gains indicate effective domain adaptation}
We report both general domain and e-commerce domain PPLR. Overall, the contextualization and domain adaptation techniques help with the PPL dropping in both cases. The jointly trained model on in-domain slot detection however clearly helps more in the e-commerce case. Moreover, since we used BERT that was finetuned on e-commerce text we again see larger gains in the domain specific testset when compared to the general domain testset (23.3\% vs 16.3\%).

\textbf{Using system dialogue acts improves intent detection}: 
From our experiments that train the TXL LMs with dialogue act information, it is clear that dialogue acts helps with relatively marginal gains in PPL (3.4\% on generic and 9.9\% on e-commerce) and WER (1.2\% Retail, 14.4\% Fastfood). When compared to other techniques we explored, we see that the impact on intent classification was higher in proportion to the gain in WER, which indicates that dialogue acts are valuable contextual signals to help with intent conveying phrases.

\textbf{Slot detection loss yields improvements on domain specific content words}: 
Rows 4 and 5 of Table \ref{table:wer_report} report the content WERR, IC and SL F1s that we obtain by incorporating the joint LM and slot detection (SD) loss. As expected, the multi-task model improves on the content words significantly (1.2\% to 4.3\% on Retail, 12.3\% to 16.3\% on Fastfood). This WER improvement also carries over to a higher SL F1 improvement, but a relatively small IC F1 improvement. This is again indicative that the improvements are mainly on recognition of in-domain slot entities and the auxiliary function words that are important to recognize intents do not benefit as much.

\textbf{Domain adaptation using BERT fusion provides maximum gains}:
Rows 6 and 7 in Table \ref{table:wer_report} illustrate the performance of the TXL LM that incorporates the BERT embedding fusion layer. Compared to the model trained with joint slot detection loss, BERT fusion model performs better on all ASR and the NLU metrics. It is evident from the results that the BERT embeddings that are derived from different user provided text helps the model effectively adapt to the domain that the embedding was derived from. The gains are amplified when complemented with the dialogue acts ability to improve on intent carrying words and the joint slot detection model leading to a WERR improving from 12.3\% to 19.2\% on the fastfood domain and 1\% to 11.8\% on the retail domain. This also carries over to an improvement on IC and SL F1 of 3.8\%, 4.3\% on retail and 2.1\%, 6.4\% on fastfood.

\section{Conclusion}

In this paper we explored different ways to robustly adapt a domain-general Transformer-XL NLM to rescore N-best hypotheses from a hybrid ASR system for task-oriented e-commerce speech conversations. We demonstrated that Transformer-XL LM trained with turn level dialogue acts benefits intent classification by improving the recognition of content words. Additionally, we show that using semantic embeddings derived from a masked language model finetuned on e-commerce domain can be effectively used to adapt a domain-general TXL LM for e-commerce domain utterance rescoring task. Finally, we introduced a new TXL training loss function to jointly predict content words along with language modeling task, this when combined with BERT fusion and dialogue acts, amplifies the WER, IC F1 and SL F1 gains. We have also shown these improvements to be statistically significant. Future work can look at integrating these methods into an end-to-end ASR system for both rescoring task and first pass LM fusion.

\bibliographystyle{acl_natbib}
\bibliography{anthology,acl2021}

\begin{thebibliography}{33}
\expandafter\ifx\csname natexlab\endcsname\relax\def\natexlab#1{#1}\fi

\bibitem[{Austin(1975)}]{john_austin:75}
John~Langshaw Austin. 1975.
\newblock How to do things with words.

\bibitem[{Brown et~al.(2020)Brown, Mann, Ryder, Subbiah, Kaplan, Dhariwal,
  Neelakantan, Shyam, Sastry, Askell, Agarwal, Herbert-Voss, Krueger, Henighan,
  Child, Ramesh, Ziegler, Wu, Winter, Hesse, Chen, Sigler, Litwin, Gray, Chess,
  Clark, Berner, McCandlish, Radford, Sutskever, and
  Amodei}]{brown2020language}
Tom~B. Brown, Benjamin Mann, Nick Ryder, Melanie Subbiah, Jared Kaplan,
  Prafulla Dhariwal, Arvind Neelakantan, Pranav Shyam, Girish Sastry, Amanda
  Askell, Sandhini Agarwal, Ariel Herbert-Voss, Gretchen Krueger, Tom Henighan,
  Rewon Child, Aditya Ramesh, Daniel~M. Ziegler, Jeffrey Wu, Clemens Winter,
  Christopher Hesse, Mark Chen, Eric Sigler, Mateusz Litwin, Scott Gray,
  Benjamin Chess, Jack Clark, Christopher Berner, Sam McCandlish, Alec Radford,
  Ilya Sutskever, and Dario Amodei. 2020.
\newblock \href {http://arxiv.org/abs/2005.14165} {Language models are few-shot
  learners}.

\bibitem[{Budzianowski et~al.(2018)Budzianowski, Wen, Tseng, Casanueva, Ultes,
  Ramadan, and Gasic}]{multiwoz:18}
Pawel Budzianowski, Tsung{-}Hsien Wen, Bo{-}Hsiang Tseng, I{\~{n}}igo
  Casanueva, Stefan Ultes, Osman Ramadan, and Milica Gasic. 2018.
\newblock \href {http://arxiv.org/abs/1810.00278} {Multiwoz - {A} large-scale
  multi-domain wizard-of-oz dataset for task-oriented dialogue modelling}.
\newblock \emph{CoRR}, abs/1810.00278.

\bibitem[{Chen et~al.(2015)Chen, Tan, Liu, Lanchantin, Wan, Gales, and
  Woodland}]{Chen:2015}
X.~Chen, T.~Tan, Xunying Liu, Pierre Lanchantin, M.~Wan, Mark J.~F. Gales, and
  Philip~C. Woodland. 2015.
\newblock Recurrent neural network language model adaptation for multi-genre
  broadcast speech recognition.
\newblock In \emph{{Interspeech} 2015, Dresden, Germany, September 6-10, 2015},
  pages 3511--3515. {ISCA}.

\bibitem[{Dai et~al.(2019)Dai, Yang, Yang, Carbonell, Le, and
  Salakhutdinov}]{dai-etal-2019-transformer}
Zihang Dai, Zhilin Yang, Yiming Yang, Jaime Carbonell, Quoc Le, and Ruslan
  Salakhutdinov. 2019.
\newblock \href {https://doi.org/10.18653/v1/P19-1285} {Transformer-{XL}:
  Attentive language models beyond a fixed-length context}.
\newblock In \emph{Proceedings of the 57th Annual Meeting of the Association
  for Computational Linguistics}, pages 2978--2988, Florence, Italy.
  Association for Computational Linguistics.

\bibitem[{Eric et~al.(2019)Eric, Goel, Paul, Sethi, Agarwal, Gao, and
  Hakkani{-}T{\"{u}}r}]{MultiWoz:19}
Mihail Eric, Rahul Goel, Shachi Paul, Abhishek Sethi, Sanchit Agarwal, Shuyang
  Gao, and Dilek Hakkani{-}T{\"{u}}r. 2019.
\newblock \href {http://arxiv.org/abs/1907.01669} {Multiwoz 2.1: Multi-domain
  dialogue state corrections and state tracking baselines}.
\newblock \emph{CoRR}, abs/1907.01669.

\bibitem[{Gandhe et~al.(2018)Gandhe, Rastrow, and Hoffmeister}]{aggandhe:18}
Ankur Gandhe, Ariya Rastrow, and Bj{\"{o}}rn Hoffmeister. 2018.
\newblock Scalable language model adaptation for spoken dialogue systems.
\newblock In \emph{{SLT Workshop} 2018, Athens, Greece}, pages 907--912.
  {IEEE}.

\bibitem[{Ghoshal and Povey(2013)}]{Ghoshal13}
Arnab Ghoshal and Daniel Povey. 2013.
\newblock Sequencediscriminative training of deep neural networks.
\newblock In \emph{in Proc. INTERSPEECH}.

\bibitem[{Irie et~al.(2018)Irie, Kumar, Nirschl, and Liao}]{irie18:radmm}
Kazuki Irie, Shankar Kumar, Michael Nirschl, and Hank Liao. 2018.
\newblock Radmm: Recurrent adaptive mixture model with applications to domain
  robust language modeling.
\newblock In \emph{IEEE International Conference on Acoustics, Speech, and
  Signal Processing}, pages 6079--6083, Calgary, Canada.

\bibitem[{Irie et~al.(2019)Irie, Zeyer, Schl{\"{u}}ter, and Ney}]{irie:19}
Kazuki Irie, Albert Zeyer, Ralf Schl{\"{u}}ter, and Hermann Ney. 2019.
\newblock Training language models for long-span cross-sentence evaluation.
\newblock In \emph{ASRU Singapore}, pages 419--426. {IEEE}.

\bibitem[{Khandelwal et~al.(2018)Khandelwal, He, Qi, and
  Jurafsky}]{khandelwal-etal-2018-sharp}
Urvashi Khandelwal, He~He, Peng Qi, and Dan Jurafsky. 2018.
\newblock \href {https://doi.org/10.18653/v1/P18-1027} {Sharp nearby, fuzzy far
  away: How neural language models use context}.
\newblock In \emph{Proceedings of the 56th Annual Meeting of the Association
  for Computational Linguistics (Volume 1: Long Papers)}, pages 284--294,
  Melbourne, Australia. Association for Computational Linguistics.

\bibitem[{Kneser and Ney(1995)}]{KN:95}
Reinhard Kneser and Hermann Ney. 1995.
\newblock Improved backing-off for m-gram language modeling.
\newblock In \emph{ICASSP}, pages 181--184. IEEE Computer Society.

\bibitem[{Li et~al.(2020)Li, Zhou, He, Wang, Yang, and
  Li}]{li-etal-2020-sentence}
Bohan Li, Hao Zhou, Junxian He, Mingxuan Wang, Yiming Yang, and Lei Li. 2020.
\newblock \href {https://doi.org/10.18653/v1/2020.emnlp-main.733} {On the
  sentence embeddings from pre-trained language models}.
\newblock In \emph{Proceedings of the 2020 Conference on Empirical Methods in
  Natural Language Processing (EMNLP)}, pages 9119--9130, Online. Association
  for Computational Linguistics.

\bibitem[{Li et~al.(2018)Li, Xu, Wang, Povey, and Khudanpur}]{keli:2018}
Ke~Li, Hainan Xu, Yiming Wang, Daniel Povey, and Sanjeev Khudanpur. 2018.
\newblock Recurrent neural network language model adaptation for conversational
  speech recognition.
\newblock In \emph{{Interspeech}, Hyderabad, India, 2-6 September 2018}, pages
  3373--3377.

\bibitem[{Liu and Lane(2016)}]{liu-lane-2016-joint}
Bing Liu and Ian Lane. 2016.
\newblock \href {https://doi.org/10.18653/v1/W16-3603} {Joint online spoken
  language understanding and language modeling with recurrent neural networks}.
\newblock In \emph{Proceedings of the 17th Annual Meeting of the Special
  Interest Group on Discourse and Dialogue}, pages 22--30, Los Angeles.
  Association for Computational Linguistics.

\bibitem[{Liu et~al.(2020)Liu, Liu, Zhang, Synnaeve, Saraf, and
  Zweig}]{Liu:2020}
Da{-}Rong Liu, Chunxi Liu, Frank Zhang, Gabriel Synnaeve, Yatharth Saraf, and
  Geoffrey Zweig. 2020.
\newblock \href {https://doi.org/10.21437/Interspeech.2020-1344}
  {Contextualizing {ASR} lattice rescoring with hybrid pointer network language
  model}.
\newblock In \emph{Interspeech 2020, 21st Annual Conference of the
  International Speech Communication Association, Virtual Event, Shanghai,
  China, 25-29 October 2020}, pages 3650--3654. {ISCA}.

\bibitem[{Maarek(2018)}]{maarek:2018}
Yoelle Maarek. 2018.
\newblock \href {https://doi.org/10.1145/3269206.3272923} {{Alexa and Her
  Shopping Journey}}.
\newblock In \emph{{Proceedings of the 27th ACM International Conference on
  Information and Knowledge Management}}, page~1. {ACM}.

\bibitem[{Mari et~al.(2020)Mari, Mandelli, and Algesheimer}]{mari:20}
Alex Mari, Andreina Mandelli, and Ren{\'{e}} Algesheimer. 2020.
\newblock \href {https://doi.org/10.1007/978-3-030-50341-3\_32} {The evolution
  of marketing in the context of voice commerce: {A} managerial perspective}.
\newblock In \emph{{HCI} in Business, Government and Organizations - 7th
  International Conference, {HCIBGO} 2020, Held as Part of the 22nd {HCI}
  International Conference, {HCII} 2020, Copenhagen, Denmark, July 19-24, 2020,
  Proceedings}, volume 12204 of \emph{Lecture Notes in Computer Science}, pages
  405--425. Springer.

\bibitem[{Mikolov and Zweig(2019)}]{mikolov:12}
Tomas Mikolov and Geoffrey Zweig. 2019.
\newblock Context dependent recurrent neural network language model.
\newblock In \emph{SLT}, pages 234--239. IEEE.

\bibitem[{Parthasarathy et~al.(2019)Parthasarathy, Gale, Chen, Polovets, and
  Chang}]{partha:19}
Sarangarajan Parthasarathy, William Gale, Xie Chen, George Polovets, and
  Shuangyu Chang. 2019.
\newblock \href {http://arxiv.org/abs/1911.04571} {Long-span language modeling
  for speech recognition}.
\newblock \emph{CoRR}, abs/1911.04571.

\bibitem[{Peskov et~al.(2019)Peskov, Clarke, Krone, Fodor, Zhang, Youssef, and
  Diab}]{multidogo:19}
Denis Peskov, Nancy Clarke, Jason Krone, Brigi Fodor, Yi~Zhang, Adel Youssef,
  and Mona Diab. 2019.
\newblock Multi-domain goal-oriented dialogues ({M}ulti{D}o{GO}): Strategies
  toward curating and annotating large scale dialogue data.
\newblock In \emph{Proc EMNLP-IJCNLP}, pages 4526--4536.

\bibitem[{Radford et~al.(2016)Radford, Metz, and Chintala}]{Radford2019}
Alec Radford, Luke Metz, and Soumith Chintala. 2016.
\newblock \href {http://arxiv.org/abs/1511.06434} {Unsupervised representation
  learning with deep convolutional generative adversarial networks}.

\bibitem[{Raju et~al.(2018)Raju, Hedayatnia, Liu, Gandhe, Khatri, Metallinou,
  Venkatesh, and Rastrow}]{anirudh:18}
Anirudh Raju, Behnam Hedayatnia, Linda Liu, Ankur Gandhe, Chandra Khatri,
  Angeliki Metallinou, Anu Venkatesh, and Ariya Rastrow. 2018.
\newblock Contextual language model adaptation for conversational agents.
\newblock In \emph{Interspeech, Hyderabad, India}, pages 3333--3337. {ISCA}.

\bibitem[{Rao et~al.(2020)Rao, Raju, Dheram, Bui, and Rastrow}]{Milind:20}
Milind Rao, Anirudh Raju, Pranav Dheram, Bach Bui, and Ariya Rastrow. 2020.
\newblock \href {https://doi.org/10.21437/Interspeech.2020-2976} {Speech to
  semantics: Improve {ASR} and {NLU} jointly via all-neural interfaces}.
\newblock In \emph{Interspeech 2020, 21st Annual Conference of the
  International Speech Communication Association, Virtual Event, Shanghai,
  China, 25-29 October 2020}, pages 876--880. {ISCA}.

\bibitem[{{Rastogi} et~al.(2019){Rastogi}, {Zang}, {Sunkara}, {Gupta}, and
  {Khaitan}}]{Rastogi:20}
Abhinav {Rastogi}, Xiaoxue {Zang}, Srinivas {Sunkara}, Raghav {Gupta}, and
  Pranav {Khaitan}. 2019.
\newblock \href {http://arxiv.org/abs/1909.05855} {{Towards Scalable
  Multi-domain Conversational Agents: The Schema-Guided Dialogue Dataset}}.
\newblock \emph{arXiv e-prints}, page arXiv:1909.05855.

\bibitem[{Reimers and Gurevych(2019)}]{reimers-gurevych-2019-sentence}
Nils Reimers and Iryna Gurevych. 2019.
\newblock \href {https://doi.org/10.18653/v1/D19-1410} {Sentence-{BERT}:
  Sentence embeddings using {S}iamese {BERT}-networks}.
\newblock In \emph{Proceedings of the 2019 Conference on Empirical Methods in
  Natural Language Processing and the 9th International Joint Conference on
  Natural Language Processing (EMNLP-IJCNLP)}, pages 3982--3992, Hong Kong,
  China. Association for Computational Linguistics.

\bibitem[{Sharma(2020)}]{sharma-2020-improving}
Arpit Sharma. 2020.
\newblock \href {https://doi.org/10.18653/v1/2020.ecnlp-1.6} {Improving intent
  classification in an {E}-commerce voice assistant by using inter-utterance
  context}.
\newblock In \emph{Proceedings of The 3rd Workshop on e-Commerce and NLP},
  pages 40--45, Seattle, WA, USA. Association for Computational Linguistics.

\bibitem[{Shenoy et~al.(2021)Shenoy, Bodapati, Sunkara, Ronanki, and
  Kirchhoff}]{shenoy2021whats}
Ashish Shenoy, Sravan Bodapati, Monica Sunkara, Srikanth Ronanki, and Katrin
  Kirchhoff. 2021.
\newblock \href {http://arxiv.org/abs/2104.11070} {Adapting long context nlm
  for asr rescoring in conversational agents}.

\bibitem[{Sun et~al.(2021)Sun, Zhang, and Woodland}]{sun2021transformer}
G.~Sun, C.~Zhang, and P.~C. Woodland. 2021.
\newblock \href {http://arxiv.org/abs/2102.06474} {Transformer language models
  with lstm-based cross-utterance information representation}.

\bibitem[{Sunkara et~al.(2020)Sunkara, Ronanki, Bekal, Bodapati, and
  Kirchhoff}]{sunkaral:2020}
Monica Sunkara, Srikanth Ronanki, Dhanush Bekal, Sravan Bodapati, and Katrin
  Kirchhoff. 2020.
\newblock Multimodal semi-supervised learning framework for punctuation
  prediction in conversational speech.
\newblock In \emph{Interspeech 2020, Shanghai, China, 25-29 October 2020},
  pages 4911--4915. {ISCA}.

\bibitem[{Vaswani et~al.(2017)Vaswani, Shazeer, Parmar, Uszkoreit, Jones,
  Gomez, Kaiser, and Polosukhin}]{vaswani:17}
Ashish Vaswani, Noam Shazeer, Niki Parmar, Jakob Uszkoreit, Llion Jones,
  Aidan~N Gomez, \L~ukasz Kaiser, and Illia Polosukhin. 2017.
\newblock Attention is all you need.
\newblock In \emph{Advances in Neural Information Processing Systems},
  volume~30, pages 5998--6008.

\bibitem[{Xiong et~al.(2018)Xiong, Wu, Zhang, and
  Stolcke}]{xiong-etal-2018-session}
Wayne Xiong, Lingfeng Wu, Jun Zhang, and Andreas Stolcke. 2018.
\newblock \href {https://doi.org/10.18653/v1/D18-1296} {Session-level language
  modeling for conversational speech}.
\newblock In \emph{Proceedings of the 2018 Conference on Empirical Methods in
  Natural Language Processing}, pages 2764--2768, Brussels, Belgium.
  Association for Computational Linguistics.

\bibitem[{Yang et~al.(2020)Yang, Liu, Gandhe, Gu, Raju, Filimonov, and
  Bulyko}]{yang:20}
Chao{-}Han~Huck Yang, Linda Liu, Ankur Gandhe, Yile Gu, Anirudh Raju, Denis
  Filimonov, and Ivan Bulyko. 2020.
\newblock \href {http://arxiv.org/abs/2011.11715} {Multi-task language modeling
  for improving speech recognition of rare words}.
\newblock \emph{CoRR}, abs/2011.11715.

\end{thebibliography}

%\appendix

\end{document}